# A theoretical revisit of giant transmission of light through a metallic nano-slit surrounded with periodic grooves

Yanxia Cui[1] and Sailing He[1, 2] *

[1] Centre for Optical and Electromagnetic Research, Zhejiang University; Joint Research Centre of Photonics of the Royal Institute of Technology (Sweden) and Zhejiang University, Zhejiang University, China

[2] Division of Electromagnetic Engineering, School of Electrical Engineering, Royal Institute of Technology, S-100 44 Stockholm, Sweden

*Corresponding author: sailing@kth.se

The giant transmission of light through a metallic nano-slit surrounded by periodic grooves on the input surface is revisited theoretically. It is shown that the influence to the transmission comes from three parts: the groove-generated surface plasmon wave (SPW), the nano-slit-generated SPW and the incident wave. The groove-generated SPW is the main factor determining the local field distribution around the nano-slit opening, which is directly related to the transmission through the nano-slit. The nano-slit-generated SPW can be considered as a disturbance to the light distribution on the input surface. The influence of the incident wave can be strongly reduced when strong surface plasmon wave is generated on the input surface by many periods of deep grooves. Our study shows that the slit-to-groove distance for a maximal transmission through the nano-slit surrounded with periodic grooves can not be predicted by several previous theories, including the magnetic field phase theory of a recent work (Phys. Rev. Lett. 99, 043902, 2007). A clear physical explanation is given for the dependence of the transmission on the slit-to-groove distance.

PACS numbers: 42.25.Bs, 42.25.Fx, 42.79.Ag, 73.20.Mf

Transmission efficiency of an isolated nano-aperture in a metallic film is quite low [1, 2]. However, by using periodic grooves around the aperture on the input surface, transmission efficiency could be greatly enhanced (see e.g. [3-6]). Such a phenomenon has been intensively studied [7-9] and can be applied in many potential applications [10, 11]. However, until now the physical mechanism of this phenomenon is still not fully explained. For example, one important issue is the dependence of the transmission on the distance between the nano-slit center and the nearest groove center, and this has been studied in Refs. [7] and [9] with inconsistent conclusions (both conclusions are not correct as one will see below in our study). In this letter, through an extensive study of the transmission (field) property for different slit-to-groove distances in various scenarios, we explain clearly the physical mechanism of extraordinary transmission through the nano-slit surrounded with periodic grooves.

Fig. 1 shows the two-dimensional (2D) schematic diagram of a metallic nano-slit surrounded by periodic grooves (on the input surface) on both sides. A TM-polarized (the magnetic field is perpendicular to the $x$-$z$ plane) plane wave impinges normally on the structure. The wavelength of the surface plasmon wave (SPW) is usually estimated by formula $\lambda_{sp} = \mathrm{Re}(\lambda_0 / \sqrt{\varepsilon_m \varepsilon_d / (\varepsilon_m + \varepsilon_d)})$, where $\varepsilon_m$ and $\varepsilon_d$ are the relative permittivities of the metal and the dielectric in immediate contact with the metal surface, respectively. The film is free-standing in the air and made of gold with permittivity $\varepsilon_m = -24.1 + i1.51$ [9] at wavelength $\lambda_0 = 800$ nm. Then we obtain $\lambda_{sp} \approx 780$ nm. We define transmission efficiency $\eta$ as the ratio of the integration of the $z$-component of the Poynting vector over the output aperture to that over the input aperture. All the numerical simulations are performed by the finite-element method (FEM) (COMSOL).

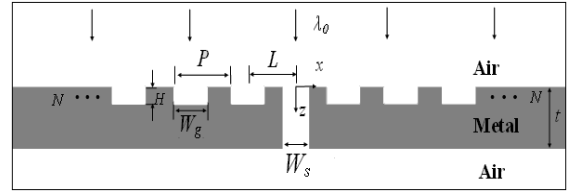

Fig. 1 Schematic diagram of the metallic nano-slit (centered at $x = 0$ with width $W_s$ milled in a metallic film of thickness $t$) surrounded with $N$ grooves (with period $P$, depth $H$ and width $W_g$) on the input air-metal surface (located at $z = 0$) on each side.

First we study the structure with groove depth $H = 60$ nm, width $W_g = 250$ nm, period $P = 720$ nm and number $N = 12$. The distance between the slit center and the nearest groove center (slit-to-groove distance) is denoted by $L$ and the width of the nano-slit is $W_s = 50$ nm. In Fig. 2(a), we show transmission efficiency $\eta$ as a function of $L$ when $t = 340$ nm (thin solid curve; forming a non-resonant nano-slit) or $t = 200$ nm (thick solid curve; resonant). Since the transmission through the nano-slit is directly related to the local field distribution around the nano-slit opening, we fill the air nano-slit with the same metal (gold) and check the magnetic field amplitude $|H_y|$ at point (0, -10 nm) as a function of distance $L$ [the dashed curve in Fig. 2(a)]. One sees that the dashed curve oscillates with the same features as the thin solid curve, and also has similar features but with some mismatches compared with the thick solid curve (e.g., the positions of the two peaks in each period have some small shifts and the relative intensities of the two peaks do not match quite well). This is because the influence of the nano-slit can be considered as a disturbance to the light distribution on the input surface. Both the grooves and the nano-slit can generate SPWs on the input surface. However, the intensity of the groove-generated SPW on the input surface is much larger than that of the nano-slit-generated SPW. Furthermore, due to destructive interference in the slit cavity, a non-resonant nano-slit gives negligible influence [2] to the local field distribution (on the input surface) generated by the grooves [this explains the good matching between the thin solid curve and the dashed curve in Fig. 2(a)]. Due to constructive interference, a resonant nano-slit will slightly influence the local field distribution generated by the grooves and further change the transmission through the nano-slit [this explains the mismatch between the thick solid curve and the dashed curve in Fig. 2(a)].

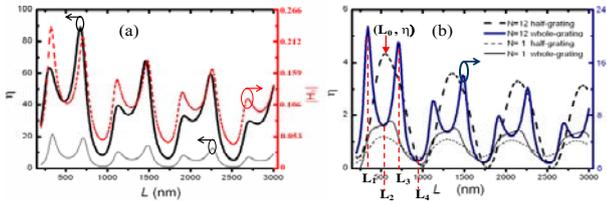

Fig. 2 (Color online) Transmission efficiency $\eta$ through a 50 nm-width nano-slit for a groove structure with $H$ = 60 nm, $W_g$ = 250 nm, $P$ = 720 nm and $N$ = 12. (a) Curves of $\eta \sim L$ when $t$ = 340 nm (thin solid, non-resonant) and $t$ = 200 nm (thick solid, resonant). Field $|H_y|$ (dashed) at point (0, -10 nm) is also shown for comparison. (b) Curves of $\eta \sim L$ when the grooves [with $N$ = 12 (thick) or $N$ = 1 (thin)] are only on the right side (dashed) and on both sides (solid) of the non-resonant nano-slit ($t$ = 340 nm).

Before studying some other features for the curves of $\eta \sim L$ in Fig. 2(a), we investigate another simple case by removing one side (left side) of the grooves (called the half-grating case hereafter). To minimize the influence of the nano-slit, we choose a non-resonant nano-slit ($t$ = 340 nm) to study. The relation of $\eta \sim L$ for the half-grating case when $N$ = 12 is shown in Fig. 2(b) by the thick dashed curve, which oscillates sinusoidally with a period of about 780 nm (i.e., $\lambda_{sp}$) and a relatively small amplitude [about a quarter of that for the thin solid curve in Fig. 2 (a)]. For example, transmission efficiency reaches a peak value of 4.3 at slit-to-groove distance $L_0$ in the first period, and its corresponding field distribution $|H_y|$ is shown in Fig. 3(a). In Fig. 3(a) one sees that the surface wave is strongly confined to the grating interface. Because the number of the grooves is not infinite, the field confinement is the strongest at the center ($x$ = 4.5 $\mu$m) of the grating and becomes much weaker at the grating edges. One can also see clearly that curve $|H_y| \sim x$ at line $z$ = -10 nm [green curve in Fig. 3(a)] decreases (from an amplitude of 0.19 at the grating center) very quickly toward $\pm x$ directions with an oscillating period of $P$ = 720 nm. However, if there is no groove on the input surface, the field is distributed as a SPW on the flat metal-air interface with a period of $\lambda_{sp} \approx$ 780 nm and a weak confinement effect [see the period (780 nm) and smaller intensity (lower than 0.1) on the left side of curve $|H_y| \sim x$ in Fig. 3(a)]. Since the nano-slit is opened in the flat metal-air interface region ($x$ < 0, larger $L$ corresponds to larger $|x|$), the thick dashed curve $\eta \sim L$ in Fig. 2(b) has the properties of period 780 nm ($\lambda_{sp}$), low amplitude (due to weak confinement), and small damping. This also indicates that groove-generated SPW influences significantly the transmission.

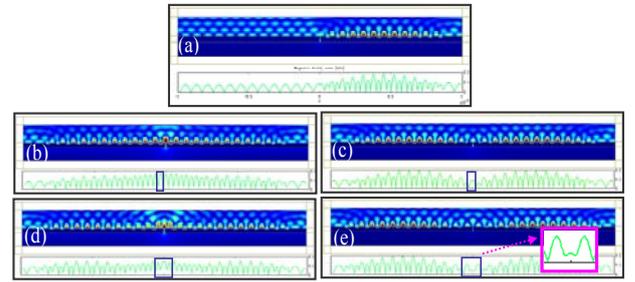

Fig. 3 (Color online) Magnetic field distributions $|H_y|$ for different situations (with $N$ = 12) considered in Fig. 2(b). (a) $L_0$ = 545 nm for the half-grating case, (b-d) $L_1$ = 340 nm, $L_2$ = 530 nm, $L_3$ = 710 nm and $L_4$ = 950 nm for the whole-grating case. The green curves show the intensity of magnetic field on horizontal line $z$ = -10 nm.

Now we come back to study the structure of Fig. 1 (called the whole-grating case hereafter). We redraw in Fig. 2(b) the relation of $\eta \sim L$ for the non-resonant nano-slit [i.e. the thin solid curve in Fig. 2(a)] by the thick solid curve, which has three main features: (a) it oscillates with a period of $\lambda_{sp} \approx$ 780 nm; (b) there are two sharp peaks in each period (e.g. $\eta$ = 21.5 at $L_1$ = 340 nm and $\eta$ = 19.5 at $L_3$ = 710 nm in the first period), whose transmission can be four times larger than that for the half-grating case ($\eta$ = 4.3). There are also two valleys in each period (e.g. $\eta$ = 6.4 at $L_2$ = 530 nm and $\eta$ = 1.2 at $L_4$ = 950 nm in the first period); (c) the peaks decrease more quickly as compared with the half-grating case. For different slit-to-groove distances the field distributions are different on both the grating interface and the flat metal-air interface (between the left and right half-gratings; opened with the nano-slit) as shown in Fig. 3(b)-(e) for $|H_y|$ distributions at the above two peaks and two valleys. For a minimal transmission situation (corresponding to slit-to-groove distance $L_2$) shown in Fig. 3(c), the field distribution on the grating interface is the same as the field distribution on the grating interface for the half-grating case in Fig. 3(a) and an oscillation peak (but not large field intensity) is formed on the flat metal-air interface [also seen clearly in the middle square region of curve $|H_y| \sim x$ in Fig. 3(c)]. Since light has been confined to the center of the left and right half-gratings, the field confinement on the flat metal-air interface (including the nano-slit area) is quite weak after a quick drop along the grating interface [see Fig. 3(c)]. For the other minimal transmission situation ($L = L_4$) shown in Fig. 3(e), the field distribution is very similar to that in Fig. 3(c), and the only difference is that there are three oscillations on the flat metal-air interface between the two half-gratings [a small one at the center and two large ones on the sides, seen from the enlarged insert for $|H_y| \sim x$ in Fig. 3(e)]. The small (or extremely small) amplitude at the center is the reason for the low transmission when $L = L_2$ (or $L = L_4$). For a maximum transmission situation ($L = L_1$) shown in Fig. 3(b), the field distribution on the grating interface is no longer the same as that for the half-grating case, instead, it is strongly confined at $x = 0$ (the opening position of the nano-slit) and an oscillation is formed between the nearest left and right grooves [see curve $|H_y| \sim x$ in the square region of Fig. 3(b), which is much stronger than that in Fig. 3(c)]. For the other maximum transmission situation ($L = L_3$) shown in Fig. 3(d), the field distribution on the grating interface is quite similar to that in Fig. 3(b) except smaller amplitude and also three oscillations on the flat metal-air interface between the two half-gratings [see curve $|H_y| \sim x$ in Fig. 3(d)]. Due to the strong field confinement on the flat metal-air interface region opened with the nano-slit, the transmission for $L = L_1$ (or $L_3$) is quite high (and the peak drops quickly as $L$ varies). In this situation, the left and right half-gratings work together as a whole grating with energy strongly concentrated around the whole-grating center (i.e. $x = 0$; note that all the bright spots in the flat metal-air interface region have higher field intensity than those on the grating interface). The flat metal-air interface between the two half-gratings supports some special interference pattern formed by two counter-propagating SPWs (of wavelength $\lambda_{sp}$) coming from the left and right half-gratings. This explains why the period of the curve $\eta \sim L$ is equal to $\lambda_{sp}$ and is not related to the groove period ($P = 720$nm). For a minimal transmission distance ($L_2$ or $L_4$), the SPWs generated by the left and right half-gratings are weakly coupled to each other and the centers of the confined energy (on both sides) are around the center of each half-grating (instead of $x = 0$). Thus, the field intensity at the nano-slit opening is much weaker than that around the center of each half-grating. In this situation, the left (right) half-grating seems to influence the transmission separately.

The authors of Ref. [9] proposed a magnetic field phase theory and argued that the maximal (or minimal) transmission through the nano-slit corresponds to constructive (or destructive) interference between the groove-generated SPW and the incident plane wave at the nano-slit center, which was realized by adjusting the slit-to-groove distance to $L \approx 0.5\lambda_{sp}$ (or $L \approx \lambda_{sp}$). According to their magnetic field phase theory, the maximal transmission should appear when $L = 0.5\lambda_{sp} + K\lambda_{sp}$ ($K$ is an integer) and there should be only one maximal transmission per period ($\lambda_{sp}$) for the curve $\eta \sim L$, which conflicts with our results shown in Fig. 2(a) that there are two transmission peaks per period. Their mistake is that they predicted the optimal distance ($L$) of maximal transmission through a nano-slit surrounded with 33 pairs of grooves by using the excitation phase of the SPW of only one groove on one side of the nano-slit. In Fig. 2(b) we plot curve $\eta \sim L$ when only one groove is milled on one side of the nano-slit (i.e., half-grating with $N = 1$) with the thin dashed curve. From this curve one sees only one oscillation peak per period, which is completely different from the thick solid curve (for the whole-grating case when $N = 12$). The theory in Ref. [9] is not reliable as one sees clearly in Fig. 2(b) that the peak position for the thin dashed curve (corresponding to a maximal transmission distance according to the theory of Ref. [9]) is around the valley of the thick solid curve (corresponding to an actually minimal transmission distance) for this special case. In addition, we plot in Fig. 2(b) curve $\eta \sim L$ (thin solid) for the whole-grating case when $N = 1$. Unlike the big difference between the thick solid curve (whole-grating) and the thick dashed curve (half-grating) in Fig. 2(b) for $N = 12$, this thin solid curve has similar features as the thin dashed curve (half-grating case with $N = 1$). This is due to lack of periodic effect of grooves in the formation of localized SPW when $N = 1$.

Another group experimentally found out in Ref. [7] that for a nano-hole surrounded with 16 nano-ring grooves, the optimal distance from the nano-hole to the center of the nearest nano-ring is $L = (K+0.25)P$, where $K$ is an integer, and $P$ is the groove period (quite close to $\lambda_{sp}$ for their experimental sample). They explained this according to their analytical model (based on a simplified structure when the nano-hole is surrounded with only one nano-ring groove) that the scattered electric field ($E_x$) at the center of the groove was $\pi/2$ out of phase with respect to the incident plane wave. They ignored the periodic effect of the grooves and made the same mistake as the authors of Ref. [9]. They did not study various experimental samples to check their conclusion. Note that the problems of 3D nano-hole and 2D nano-slit surrounded with grooves have similar properties due to similar physical mechanism. Our simulations show that for different groove structures the smallest slit-to-groove distance for maximal transmission varies quite much, instead of being fixed at $0.5\lambda_{sp}$ or $0.25\lambda_{sp}$ (or $0.25P$). In Fig. 4, the thin curves show the normalized transmission efficiency ($\eta_n$) of light through the nano-slit as a function of distance $L$ for four different sets of groove depth and width (with the same groove period). One can see clearly that the shapes of thin curves $\eta_n \sim L$ (including the positions of the peaks and valleys) are completely different (for the same $P$ and same $\lambda_{sp}$), which is attributed to different local field distributions.

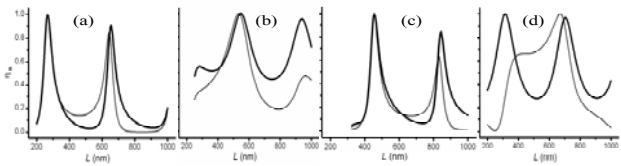

Fig. 4 Normalized transmission efficiency as $L$ varies when the incident wave is a plane wave (thin curves) or two point sources at (±11 μm, -10 nm) (thick curves). Here $W_s$ = 50 nm, $t$ = 340 nm (non-resonant case), $P$ = 780 nm and $N$ = 12. The groove depth and width are (a) $H$ = 60 nm, $W_g$ = 250 nm; (b) $H$ = 60 nm, $W_g$ = 250 nm; (c) $H$ = 60 nm, $W_g$ = 250 nm; (d) $H$ = 10 nm, $W_g$ = 250 nm.

Next we check whether the directly incident plane wave plays a significant role in the transmission enhancement or suppression of light through the nano-slit (the authors of both Refs. [7] and [9] believe that the directly incident plane wave plays a significant role). We replace the plane wave with two separated point sources (actually line sources in a 3D space) located at (±11 μm, -10 nm). In Fig. 4, we show curves $\eta_n \sim L$ for four different sets of groove depth and width for this side-point-source case (thick curves) and compare with the incident plane wave case (thin curves). When the groove depth is large ($H$ = 60 nm), one can see in Fig. 4(a)-(c) that the shapes of curves $\eta_n \sim L$ for the two different source cases are quite similar, in particular, the positions of the transmission peaks are quite close to each other, although some small mismatches exist. However, when the groove depth is small ($H$ = 10 nm), curves $\eta_n \sim L$ for the two different source cases have completely different shapes, as shown in Fig. 4(d). This can be explained as follows. For the side-point-source case, there is little light illuminating directly on the nano-slit, hence curve $\eta_n \sim L$ is mainly determined by the groove-generated SPW. When a plane wave impinges directly onto the slit opening, it will interfere with the groove-generated SPW [note that all the magnetic fields are in the same direction ($y$-direction)]. For periodic grooves with a large depth, the intensity of the groove-generated SPW near the input surface is very strong, usually several times larger than the incident field. Thus the influence of the incident field to light distribution (on the input surface) generated by the grooves is negligible, and consequently the curves of $\eta_n \sim L$ for the side-point-source case and the plane wave incident case are quite similar. However, for shallow grooves, the input surface can be approximated as a flat metallic surface and the groove-generated SPW is weak. After interfering at the slit opening with this weak SPW generated by the grooves, the incident plane wave changes quite much the shape of $\eta_n \sim L$ as compared with the incident wave from the two point sources.

In summary, through the above analysis we have shown the influence to transmission through the nano-slit surrounded with periodic grooves should come from three parts: the groove-generated SPW, the nano-slit-generated SPW, and the incident wave. We have also shown that several previous theories, including the magnetic field phase theory of Ref. [9], can not be used to predict the slit-to-groove distance for a maximal transmission through the nano-slit. We have given a clear physical explanation for the dependence of transmission on the distance between the nano-slit and the nearest groove by studying and comparing various scenarios. The authors would like to thank Prof. Yoichi Okuno, Dr. Nicholas Fang and Zhechao Wang for helpful discussions.

[1] T. Thio, K. M. Pellerin, R. A. Linke, H. J. Lezec and T.W. Ebbesen, Opt. Lett. 26, 1972 (2001).
[2] Y. Xie, A. R. Zakharian, J. V. Moloney and M. Mansuripur, Opt. Express 12, 6106 (2004).
[3] F. J. Garcı́a-Vidal, H. J. Lezec, T.W. Ebbesen and L. Martin-Moreno, Phys. Rev. Lett. 90, 213901 (2003).
[4] H. J. Lezec, A. Degiron, E. Devaux, R. A. Linke, L. Martin-Moreno, F. J. Garcia-Vidal and T. W. Ebbesen, Science 297, 820 (2002).
[5] T. Thio, H. J. Lezec, T. W. Ebbesen, K. M. Pellerin, G. D. Lewen, A. Nahata and R. A. Linke, Nanotechnology 13, 429 (2002).
[6] P. Lalanne and J. P. Hugonin, Nat Phys 2(8), 551 (2006).
[7] H. J. Lezec and T. Thio, Opt. Express 12, 3629 (2004).
[8] A. Degiron and T.W. Ebbesen, Opt. Express 12(16), 3694 (2004).
[9] O. T. A. Janssen, H. P. Urbach and G.W.'t Hooft, Phys. Rev. Lett. 99, 043902 (2007).
[10] M. J. Levene, J. Korlach, S. W. Turner, M. Foquet, H. G. Craighead and W. W. Webb, Science 299, 682 (2003)
[11] T. Ishi, J. Fujikata, K. Makita, T. Baba and K. Ohashi, Jpn J. Appl. Phys. 44, L364 (2005).